\documentclass[fleqn,twoside]{article}
\usepackage{espcrc2-mod}

\usepackage{graphicx}
\usepackage[figuresright]{rotating}

\newcommand{\AmS}{{\protect\the\textfont2
  A\kern-.1667em\lower.5ex\hbox{M}\kern-.125emS}}
\def\beq{\begin{equation}}
\def\eeq{\end{equation}}

\def\beqn{\begin{eqnarray}}
\def\eeqn{\end{eqnarray}}
\def \ep{\epsilon}
\def \vep{\varepsilon}
\def \as{\alpha_{\rm s}}
\def\VEV#1{\left\langle #1\right\rangle}
\def\abs#1{\left|#1\right|}

\newcommand\sss{\scriptscriptstyle\rm}
\newcommand\xsec{\frac{d\sigma}{dx}}

\newcommand\xsecO{\frac{d\sigma}{dO}}

\newcommand\xsborn{\left(\xsec\right)_{\sss B}}
\newcommand\xsvirt{\left(\xsec\right)_{\sss V}}
\newcommand\xsreal{\left(\xsec\right)_{\sss R}}

\newcommand\IMC{I_{\sss MC}}

\newcommand\MCatNLO{{\rm MC}@{\rm NLO}}

\newcommand\stepf{\Theta}

\newcommand\kt{k_{\sss {\rm T}}}
\newcommand\pt{p_{\sss {\rm T}}}

\newcommand\epem{e^+e^-}
\newcommand\ptb{\hat{p}_{\sss T}}
\newcommand\ptbN{\hat{p}_{\sss T}^N}
\newcommand\ptB{p_{\sss T}}
\newcommand\ptBN{p_{\sss T}^N}
\newcommand\Tobs{{\cal T}}
\newcommand\pshift{{\cal P}}
\newcommand\nE{n_{\sss E}}
\newcommand\xH{x_{\sss H}}
\newcommand\mH{m_{\sss H}}

\title{QCD at high energy~\footnotemark}

\author{Stefano Frixione \address{LAPTH, Annecy, France, and CERN, 
                     TH Division, Geneve, Switzerland}\footnotemark}
                                        
\begin{document}

\begin{abstract}
I review recent results in QCD at high energy, emphasizing the role 
of higher-order computations, power corrections, and Monte Carlo
simulations in the study of a few discrepancies between data and 
perturbative predictions, and discussing future prospects.
\vspace{1pc}
\end{abstract}

\maketitle

\footnotetext[1]{Plenary talk at XXXI International
Conference on High Energy Physics (ICHEP), Amsterdam, July 2002.}
\footnotetext[2]{On leave of absence from INFN, Sez. di Genova, Italy.}

\section{INTRODUCTION}
After more than 25 years of considerable theoretical and experimental
efforts, it appears that QCD is the theory of strong interactions.
Ideally, in high-energy QCD one needs one single piece of information
from the experiments: the value of $\as$.  Starting from that measured
value, every observable can be computed from first principles. In
practice this is not feasible, since we don't know how to perform
calculations in terms of the hadrons that experiments measure in their
detectors. Perturbation theory offers a viable way out, since it
allows to prove, at least formally, the so-called factorization
theorems. These give explicit prescriptions to write physical
observables as the convolution of short- and long-distance parts, up
to terms suppressed by the power of some large scale. We can imagine
this factorization to occur at an arbitrary scale $\mu$; with a
suitable choice of $\mu$ the short-distance pieces, which are entirely
expressed in terms of quarks and gluons, are perturbatively
calculable. The long-distance pieces (such as parton densities) cannot
be computed in perturbation theory, but their dependence on $\mu$ can.
Furthermore, they are universal, which means that they don't depend on
short-distance physics, but solely on the nature of the hadrons
involved, which is a key factor for perturbative QCD to have
predictive power.

Pending a general solution of QCD, the computing framework 
based on perturbation theory may be regarded as a hypothesis, which
needs to be supported, or disproved, by experimental observations.
Countless tests have indeed been successful, convincing us of the
correctness of this approach and of the capability of QCD to
describe strong interactions; in many areas precise measurements,
rather than tests, are being carried out. This success may give 
to the non-expert the impression that current efforts in theoretical QCD
are perhaps technically appealing, but not compelling physics-wise. 
To counter this view, it is worth reminding that in the past decade 
the studies of several technically-involved problems, such as computations 
to next-to-leading order accuracy, resummation techniques, and Monte Carlo
simulations, have been key factors to the outstanding achievements of LEP, 
SLC, HERA, and Tevatron. However, the solutions devised so far are not 
sufficient any longer. For an improvement of the accuracy in the 
extraction of $\as$, for a deeper understanding of the interplay between 
perturbative and non-perturbative physics, and for a realistic modelling of
Tevatron Run II and LHC physics, new ideas and computations are
necessary. It must be clear that such investigations are not only
relevant to the study of QCD itself, but also to a variety of other
issues, from SM precision tests to searches of beyond-the-SM
physics. Besides, a few unsatisfactory results remain in QCD,
which deserve further studies.

Needless to say, this review cannot be complete, and I'll have to
leave out several interesting results, such as new NLO calculations,
progress in small-$x$ physics, diffraction, fully numerical
computations, and spin physics. Also, I'll not present most of the
recent experimental results in high-energy QCD, since they can be
found in K.~Long's writeup \cite{KLong}. I'd rather use a few
phenomenological examples to discuss some theoretical advancements,
and open problems. Related topics can be found in Z.~Bern's writeup
\cite{ZBern}. I'll quote papers submitted to this conference as
[S-NNN], S and NNN being the session and paper numbers respectively.

\section{HEAVY FLAVOURS}
Heavy flavour production is one of the most extensively studied topics
in QCD. An impressive amount of data is available, for basically all
kinds of colliding particles. The non-vanishing quark mass allows the
definition of open-heavy-quark cross sections (whereas for light
quarks one must convolute the short-distance cross sections with
fragmentation functions, in order to cancel final-state collinear
divergences).  On the other hand, the presence of the mass makes the
calculation of the matrix elements more involved. The breakthrough was
the computation of total $Q\bar{Q}$ hadroproduction rates to NLO
accuracy \cite{Nason:1987xz,Beenakker:1990ma}, readily extended to
other production processes and more exclusive
final states \cite{Ellis:1988sb,Smith:1991pw,Drees:1992eh,Nason:1989zy,%
Mangano:jk,Frixione:1993dg,Laenen:1992zk,Laenen:1992xs}. 
The resummation of various classes of large logarithms affecting these 
fixed-order computations, such as threshold, large-$\pt$, and small-$x$ 
logs, has also been accomplished, typically at the 
next-to-leading log (NLL) accuracy.

In those kinematical regions not affected by large logs, the mass of
the heavy quark sets the hard scale. Furthermore, the impact of
effects of non-perturbative origin (such as colour drag or intrinsic
$\kt$) is known to be larger the smaller the quark mass and CM energy.
Thus, top physics is expected to be the ideal testing ground for
perturbative computations.
\begin{figure}[htb]
\includegraphics[width=15pc]{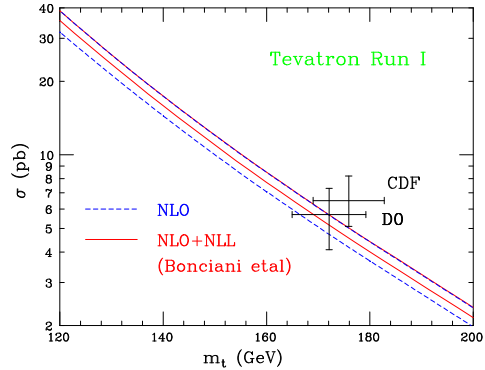}
\vskip -2.2pc
\caption{Total $t\bar{t}$ rate at the Tevatron.}
\label{fig:topatTEV}
\end{figure}
The agreement between NLO results \cite{Nason:1987xz} (dashed lines
-- the band is the spread of the prediction due to scale variation), 
and Tevatron Run I data \cite{Affolder:2001wd,Abazov:2002gy}, shown in
fig.~\ref{fig:topatTEV} for total $t\bar{t}$ rates, appears in fact
to be satisfactory. The inclusion of soft-gluon effects (solid 
lines), resummed to NLL accuracy according to the computation of 
ref. \cite{Bonciani:1998vc}, is seen to increase only marginally the NLO 
prediction, while sizably reducing the scale uncertainty. Top production 
appears therefore under perturbative control. More stringent tests will
be performed in Run II: the errors on mass and rate will be smaller,
and measurements will be performed of more exclusive $t\bar{t}$ observables 
and of single-top cross section (for which fully differential NLO results 
are now available \cite{Harris:2002md}).

Bottom quarks are copiously produced 
at colliders, and precise data for single-inclusive distributions
have been available for a long time. It is well known (see
ref. \cite{Frixione:1997ma} for a review) that NLO predictions
are about a factor of two smaller than data at the SpS and at 
the Tevatron (on the other hand, the shape of the $\pt$ spectrum of 
the centrally-produced $b$ is fairly well described by QCD).
In a recently-published measurement \cite{Acosta:2001rz} of the
$B^+$ $\pt$ spectrum, CDF find that the average data/theory ratio 
is $2.9\pm 0.2\pm 0.4$. However, this worrisome result is largely 
due to an improper computation of the NLO cross section. Let me remind 
that the spectrum of a $b$-flavoured meson $B$ is computed as follows:
\beq
\frac{d\sigma_B}{d\ptB}=\int dzd\ptb D(z;\ep)
\frac{d\sigma_b}{d\ptb}\delta(\ptB-z\ptb),
\label{btoBsec}
\eeq
where $\ptb$ ($\ptB$) is the transverse momentum of $b$ ($B$),
$d\sigma_b$ is the cross section for open-$b$ production, and
$D(z;\ep)$ is the non-perturbative fragmentation function (NPFF), 
which describes the $b\to B$ fragmentation. NPFF is not calculable
from first principles, and the free parameter it contains ($\ep$)
is fitted to data after assuming a functional form in $z$
(such as Peterson \cite{Peterson:1982ak}, Kartvelishvili
\cite{Kartvelishvili:1977pi}, etc). This fit is typically performed
using eq.~(\ref{btoBsec}), identifying the l.h.s. with $\epem$
data. It follows that the value of $\ep$ is strictly correlated to the
short-distance cross section $d\sigma_b$ used in the fitting
procedure, and thus is {\em non-physical}. When eq.~(\ref{btoBsec}) is
used to predict $B$-meson cross sections, it is therefore mandatory to
make consistent choices for $\ep$ and $d\sigma_b$.  This has not been
done in ref. \cite{Acosta:2001rz}: for $d\sigma_b$, the NLO result of
ref. \cite{Nason:1989zy} is used, but the value of $\ep$ adopted
(0.006) has been derived in the context of a LO, rather than NLO,
computation. On the other hand, if a more appropriate value of $\ep$
is chosen ($\sim$0.002 \cite{Nason:1999zj}), the theoretical
prediction increases by a mere 20\% \cite{Cacciari:2002pa}, still
rather far from the data.

There are, however, a couple of observations which save the day.
First, one has to remark that in the upper end of the $\pt$ range
probed by CDF ($\pt\sim 20$~GeV), large-$\log\pt/m$ effects may
start to show up. Therefore, FONLL computations \cite{Cacciari:1998it}
should be used rather than NLO ones. The FONLL formalism
consistently combines (i.e., avoids overcounting) the NLO result
with the cross section in which $\pt/m$ logs are resummed
to NLL accuracy (such resummed cross section is sometimes 
incorrectly referred to as ``massless''). Thus, FONLL can describe
both the small-$\pt$ ($\pt\sim m$, where resummed results don't make sense)
and the large-$\pt$ ($\pt\gg m$, where NLO results are not reliable)
regimes. The second observation concerns
again the NPFF: $d\sigma_b/d\ptb$ is a rather steeply falling 
function, and one can approximate it with $C/\ptbN$ in the
whole $\ptb$ range; then (from eq.~(\ref{btoBsec}))
$d\sigma_B/d\ptB=D_NC/\ptBN$, where $D_N=\int dz z^{N-1} D(z)$
is the $N^{th}$ Mellin moment of the NPFF. This fact has been noticed
some time ago \cite{Frixione:1997ma}, and $D_NC/\ptBN$ is seen to 
approximate the exact result fairly well \cite{Nason:1999ta}. Since $N=3$--5
(at the Tevatron), it follows that, in order to have an accurate 
prediction for the $\pt$ spectrum in hadroproduction, it is mandatory 
that the first few Mellin moments computed with $D(z)$ agreed 
with those measured. In ref. \cite{Cacciari:2002pa},
it is pointed out that this is not the case, in spite of the fact that 
the prediction for the inclusive $b$ cross section in $\epem$ collisions,
obtained with {\em the same} $D(z)$, displays an excellent agreement
with the data. There may seem to be a contradiction in this
statement: if the shape is reproduced well, why this is not true for Mellin
moments? The reason is that when fitting $D(z)$ one excludes the region
of large $z$, since it is affected by Sudakov logs, and by complex 
non-perturbative effects which are unlikely to be described by the
NPFF. On the other hand, the large-$z$ region is important for the
computation of $D_N$ (because of the factor $z^{N-1}$ in the
integrand). Therefore, for the purpose of predicting $B$-meson
spectra at colliders, ref. \cite{Cacciari:2002pa} advocates the
procedure of fitting the NPFF directly in the $N$-space. A fit to the
second moment (denoted as $N=2$ fit henceforth) is found to fit well
all the $D_N$'s for $N$ up to 10; and although Kartvelishvili's form
is used, Peterson's gives comparable results.
\begin{figure}[htb]
\includegraphics[width=15pc]{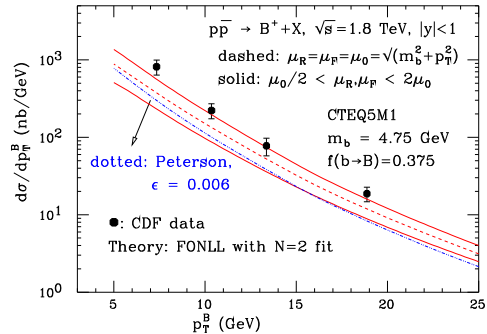}
\vskip -2.2pc
\caption{$B^+$ data vs theory \cite{Cacciari:2002pa}.}
\label{fig:BplusCN}
\end{figure}
Using the FONLL computation, and a $N=2$ fit for the NPFF, the average 
data/theory ratio reduces to $1.7\pm 0.5\pm 0.5$ \cite{Cacciari:2002pa}. 
Taking the scale uncertainty into account, $B^+$ data appear to be
compatible with QCD predictions (see fig.~\ref{fig:BplusCN}).

If one wants to avoid the pitfalls of NPFF's, an alternative possibility
consists in considering $b$-jets rather than $B$ mesons, since in this
case the NPFF simply doesn't enter the cross section. 
The comparison between NLO predictions for $b$-jets \cite{Frixione:1996nh}
and D0 measurements \cite{Abbott:2000iv} is indeed satisfactory: data are 
consistent with theory in the range $25<E_{\sss T}^{b-jet}<100$~GeV.
Overall, one can conclude that $b$ data at the Tevatron are reasonably 
described by NLO QCD. It is worth mentioning that some existing results, 
presented in terms of $b$-quark cross sections, are likely affected by
the findings of ref. \cite{Cacciari:2002pa}, and need to be reconsidered.
Among the various mechanisms which can further increase the theoretical 
predictions, small-$x$ \cite{Catani:1990eg,Collins:1991ty} and threshold
resummations \cite{Bonciani:1998vc} will probably play a secondary
role wrt NNLO contributions, which are expected to be large given
the size of the K-factor at the NLO.

\begin{figure}[htb]
\includegraphics[width=15pc]
{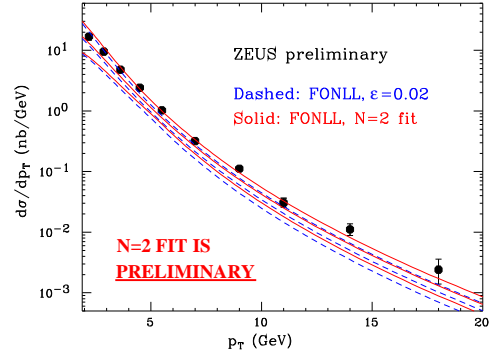}
\vskip -2.2pc
\caption{ZEUS $D^*$ data in $\gamma p$ collisions vs FONLL predictions}
\label{fig:ZEUSDstar}
\end{figure}

I now turn to the case of charm production.
A thorough discussion on this topic is beyond the scope
of this review, and I'll only give the briefest of the summaries
(which will not do any justice to the field). LEP data for total rates 
are in agreement with NLO QCD predictions \cite{Drees:1992eh};
the shapes of single-inclusive $D^*$ spectra in $\gamma\gamma$ collisions 
are as predicted by NLO QCD \cite{Frixione:1999if}, whereas normalization
is off by a factor 1.5--2, but still consistent with QCD when theoretical
uncertainties are taken into account. The vast majority of fixed target
hadro- and photoproduction data are well described by NLO computations,
but only if predictions for single-inclusive distributions and correlations
are supplemented by some parametrizations of non-perturbative phenomena 
(such as intrinsic $\kt$). At HERA, DIS data are in agreement with
NLO QCD results \cite{Laenen:1992xs}.  In photoproduction, some
concerns have arisen in the past because of the discrepancy
between ZEUS and H1 measurements in the comparison with theory:
H1 \cite{Adloff:1998vb} appears to be in perfect agreement with QCD,
whereas ZEUS \cite{Breitweg:1998yt} is at places (for intermediate
$\pt$'s and large $\eta$'s) incompatible with NLO predictions. ZEUS
have submitted to this conference [5-786] data with unprecedented
coverage at large $\pt$. The comparison to FONLL
predictions \cite{Cacciari:2001td,Frixione:2002zv}, shown in
fig.~\ref{fig:ZEUSDstar}, appears to be satisfactory, although data
are marginally harder than theory. The agreement improves if a $N=2$
fit for the NPFF is adopted (this result is still preliminary).

\begin{figure}[htb]
\includegraphics[width=15pc]
{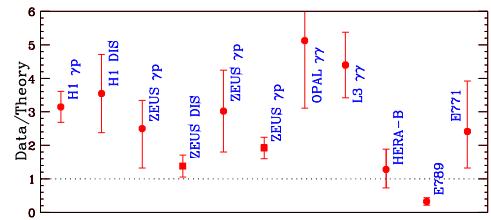}
\vskip -2.2pc
\caption{Ratios data/theory for $b$ rates.}
\label{fig:brates}
\end{figure}

Let me finally mention the increasing amount of measurements for
$b$ rates from fixed-target \cite{Abt:2002rd}, 
HERA \cite{Adloff:1999nr,Breitweg:2000nz},
[5-783,5-784,5-785,5-1013,5-1014], and 
LEP \cite{Acciarri:2000kd}, [5-366,5-475]
experiments. A summary
of the situation, in the form of ratios data/NLO QCD, is presented
in fig.~\ref{fig:brates}. While the fixed-target measurements are
in overall agreement with QCD, HERA and LEP measurements
are largely incompatible with theory. I find this hard to reconcile 
with the results presented so far, and the size of the discrepancy
also makes it problematic to find an explanation in terms of 
beyond-the-SM physics, let alone higher orders in QCD.
It is necessary to note that in many cases the experimental results are 
extrapolated to the full phase space from a rather narrow visible region.
It is encouraging to note that in the cases of the recent ZEUS measurements 
in DIS~[5-783] and photoproduction~[5-785] (full boxes in 
fig.~\ref{fig:brates}), results for single-inclusive distributions are 
given too, which are seen to be fully compatible with the corresponding NLO
predictions.

\section{POWER CORRECTIONS}
The necessity of understanding long-distance effects in final-state
measurements is not peculiar to $b$ hadroproduction. The hadron-parton
duality assumption states that there is a class of observables (such
as jet variables or event shapes in $\epem$ collisions) whose description 
in terms of quarks and gluons is expected to reproduce the data, up to
terms suppressed by some inverse power of the hard scale $Q$ of the
process (power corrections). These terms are usually estimated by
comparing the parton- and hadron-level predictions of Monte Carlo (MC)
generators. This procedure is not really satisfactory, since MC
parameters are tuned to data (which creates a bias on the
``predictions'' for power-suppressed effects), and since the
definition of parton- and hadron-level is far from being
straightforward.

It is remarkable that we can get insight on non-perturbative physics 
from perturbative considerations. The perturbative series, being asymptotic, 
can be summed to all orders only after defining a summation procedure (in a 
rather arbitrary manner); one assumes that this technical manipulation 
mimics the role played in Nature by non-perturbative effects, which are
necessary for QCD to be self-consistent. The summation procedure must 
eliminate the divergence of the perturbative series, but some 
finite quantities are left unconstrained.
Thus, the idea is to use the ambiguities of the summation procedure 
to study non-perturbative effects. Although the regularization of
the divergence can be technically very complicated, it can always 
be seen as a prescription to deal with the Landau pole of $\as$. 
The idea of ref. \cite{Dokshitzer:1995zt,Dokshitzer:1995qm} 
(DMW from now on) is to bypass such a prescription by {\em defining} 
$\as$ in the infrared (IR) region, assuming its universality; thus,
$\as$ should effectively measure confinement effects in inclusive 
quantities (in order to use such an $\as$ in actual computations,
it is also necessary to assume that the concepts of quarks and gluons
still make sense in the IR). After giving the gluon a
fake (``trigger'') mass $\mu$, the (fully inclusive) observable 
under study is computed in perturbation theory; the small-$\mu$ behaviour 
determines the power $p$ of the leading power-suppressed term
$A_p/Q^p$. The coefficient $A_p$ cannot be 
computed, but can be expressed as an integral of $k^{p-1}\as(k)$ over 
$0<k<\mu_{\sss I}$, with $\mu_{\sss I}\sim {\cal O}({\rm GeV})$.
Since all the (non computable) power-correction effects are contained
in this coefficient, one effectively gets a factorization formula.
A class of observables of great physical relevance is that of event
shapes for which $p=1$; their mean values can be computed with the 
DMW approach. The property of $\as$ universality is {\em formally}
even more far-reaching; for example, it implies that the non-perturbative 
effects it describes exponentiate for observables that do
so \cite{Korchemsky:1994is,Korchemsky:1995zm,Dokshitzer:1997ew}. This
offers the possibility of studying not only mean values, but 
distributions. The results of DMW for mean values can then 
be recovered by an expansion of the (Sudakov) exponent and subsequent
average. More precisely, if $\Tobs$ denotes the observable, the 
Sudakov is expanded in the region $\mu_{\sss I}/Q\ll\Tobs\ll 1$, 
and only the first non-trivial term in the expansion is kept.

\begin{figure}[htb]
\includegraphics[width=15pc]{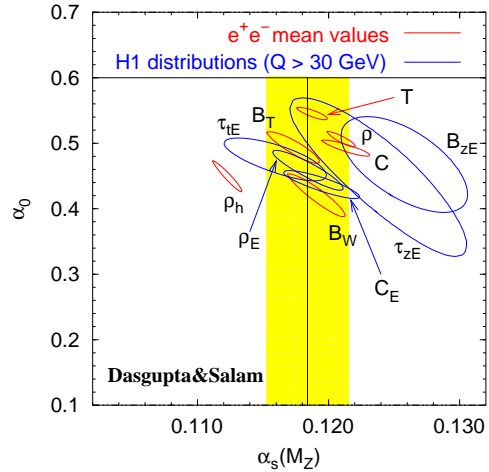}
\vskip -2.2pc
\caption{Results for $\as(M_Z)$ and $\alpha_0$ from DIS and $\epem$
event shapes. From \cite{Salam:2002vj}.}
\label{fig:asazDIS}
\end{figure}

In this context, factorization derives from the hypothesis of $\as$ IR 
universality, which is an effective description of long-distance effects, 
and as such must be insensitive to the details of parton dynamics. For
example, these details are irrelevant when one sums inclusively over 
the decay products of any parton branchings. It has been
observed \cite{Nason:1995hd} that inclusiveness is lost to a certain
extent when recoil effects (i.e., higher orders) are considered, and 
thus factorization breaks down in this case. One may still insist 
that factorization holds, and devise a procedure to systematically
account for those effects which would break it in a naive
treatment \cite{Dokshitzer:1997iz,Dokshitzer:1998pt,Dasgupta:1998xt}.
The results for mean values and distributions of event shapes can
be written as follows:
\beqn
&&\!\!\!\!\!\!\!\!\langle \Tobs\rangle \!=\! 
\langle \Tobs\rangle_{pert} + c_{\Tobs}\pshift ,
\label{EVmean}
\\
&&\!\!\!\!\!\!\!\!\frac{d\sigma}{d\Tobs}(\Tobs)\!=\! 
\left.\frac{d\sigma}{d\Tobs}\right|_{pert}(\Tobs-c_{\Tobs}\pshift),
\label{EVdistr}
\\
&&\!\!\!\!\!\!\!\!\pshift\!=\!\frac{4C_F}{\pi^2}{{\cal M}}
\frac{\mu_{\sss I}}{Q}
\Big[{\alpha_0}(\mu_{\sss I})-\as(Q)+{\cal O}(\as^2(Q))\Big],
\phantom{aaaa}
\label{pshift}
\\
&&\!\!\!\!\!\!\!\!\mu_{\sss I}{\alpha_0}(\mu_{\sss I})\!=\!
\int_0^{\mu_{\sss I}}dk\as(k).
\eeqn
Here, $c_{\Tobs}$ is a computable coefficient,
``pert'' means perturbatively-computed, and ${\cal M}$ includes the two-loop 
results of refs. \cite{Dokshitzer:1997iz,Dokshitzer:1998pt,Dasgupta:1998xt}; 
in principle, ${\cal M}$ can depend upon $\Tobs$, but (accidentally) it 
does not. The $\as(Q)$ and $\as^2(Q)$ terms 
in eq.~(\ref{pshift}) are there to avoid double
counting with the perturbatively-computed part, which shows again that
long- and short-distance effects are correlated. Since no small
parameter is involved in the two-loop computation of ${\cal M}$, one may
wonder whether factorization could be spoiled beyond two loops. Although 
it is {\em argued} that this is not the case \cite{Dokshitzer:1997iz}, a
comparison with the data is mandatory. Also notice that the formulae
above need to be modified when more complicated kinematical effects
have to be described, as in the case of broadenings \cite{Dokshitzer:1998qp}.
Eqs.~(\ref{EVmean})--(\ref{pshift}) are used to fit the data in terms
of $\as(Q)$ and $\alpha_0 (\mu_{\sss I})$. Updated analyses relevant
to $\epem$ collisions have been presented to this conference 
[5-228,5-229,5-389], \cite{MovillaFernandez:2002hu} 
-- see also \cite{MovillaFernandez:2001ed}. Results 
obtained from mean values are satisfactory, with comparable $\alpha_0$'s
obtained from different observables, and $\as(M_Z)$ values fairly
consistent with the world average. The situation worsens in the case of
distributions: $\alpha_0$ universality holds at $\sim$25\% level
(1--2 $\sigma$), and $\as(M_Z)$ values are systematically lower than 
the world average, especially for observables such as wide broadening
$B_W$ and heavy jet mass $M_H$. This fact is disturbing
since the same data for distributions, with hadronization effects
described by MC's, return $\as(M_Z)$ values in much better agreement 
with the world average.
\begin{figure}[htb]
\includegraphics[width=15pc]{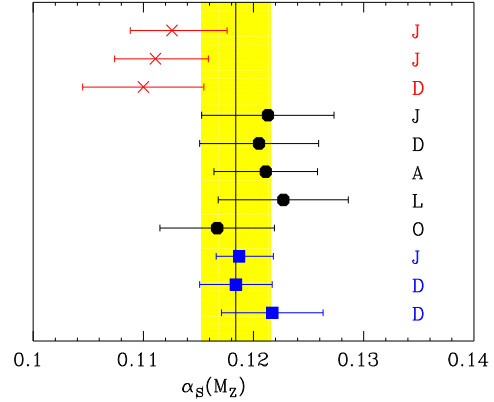}
\vskip -2.2pc
\caption{Results for $\as(M_Z)$ from event shapes in $\epem$ collisions.}
\label{fig:asres}
\end{figure}
These findings are summarized in fig.~\ref{fig:asres}, where the results 
obtained with the DMW approach are shown as boxes and crosses for
mean values and distributions respectively; the world average
$\as(M_Z)=0.1184\pm 0.0031$ \cite{Bethke:2001ih} is also shown.
Data have been taken in
the PETRA, LEP and LEP2 energy range, and analyses have been 
presented by Aleph [5-296], Delphi [5-228,5-229], Jade [5-389]
\cite{MovillaFernandez:2001ed}, L3 [5-495], and Opal [5-368].
One of the two results presented in ref. \cite{MovillaFernandez:2001ed},
reported as the uppermost cross in fig.~\ref{fig:asres}, is obtained
by excluding $B_W$ from the fit. Recently, the NLL resummation
of many DIS event shapes has been achieved (\cite{Dasgupta:2002dc}, and 
references therein). Power-correction effects can then be studied
similarly to what done for $\epem$ collisions, and the results 
are intriguing. Fig.~\ref{fig:asazDIS} (taken from ref. \cite{Salam:2002vj})
presents the $\as(M_Z)$ and $\alpha_0$ values obtained from fitting 
event shape {\em distributions} in DIS, and event shape {\em means} in
$\epem$. The DIS and $\epem$ results are largely consistent, which 
implies, in view of what shown in fig.~\ref{fig:asres}, that event 
shape distributions in $\epem$ and DIS prefer different values 
for $\as(M_Z)$.

Although not compelling from the statistical point of view, these 
results for event shape distributions in the DMW approach 
may hint to the necessity of a more complete description
of hadronization effects. As mentioned before,
eq.~(\ref{EVdistr}) results from keeping the first non-trivial term
in a Taylor expansion. If no expansion is made, from rather general
factorization arguments in the two-jet limit $\Tobs\to 0$
one gets the following formula \cite{Korchemsky:1999kt}
\beq
\frac{d\sigma}{d\Tobs}(\Tobs)=
\int_0^{\Tobs Q} d\vep f_{\Tobs}(\vep)
\left.\frac{d\sigma}{d\Tobs}\right|_{pert}
\left(\Tobs-\vep/Q\right),
\label{EVshape}
\eeq
where $f_{\Tobs}(\vep)$ is known as {\em shape function}. 
DMW formulae are recovered with 
$f_{\Tobs}(\vep)=\delta(\vep-Qc_{\Tobs}{\cal P})$; in the general case,
the first Mellin moment of $f_{\Tobs}$ has the same meaning as $\alpha_0$ 
of DMW. With eq.~(\ref{EVshape}) it is not necessary to assume that
$\Tobs Q\gg\mu_{\sss I}$ (all terms $1/(\Tobs Q)^n$ are now expected
to be included), and therefore the fit ranges can be extended.
Similarly to $\alpha_0$, $f_{\Tobs}$ cannot be computed from first
principles; thus, in order not to lose predictive power, a 
functional form depending on a small number of parameters must be 
assumed \cite{Korchemsky:2000kp,Gardi:2001ny,Gardi:2002bg}, keeping in
mind that QCD dynamics and Lorentz invariance considerations
\cite{Belitsky:2001ij} can be used to severely constrain the form of a
more general shape function, from which $f_{\Tobs}$ is derived,
independently of phenomenological arguments. In a different approach
(DGE \cite{Gardi:2001ny,Gardi:2001di}), whose final result has the
same form as eq.~(\ref{EVshape}), it is suggested to combine Sudakov
and renormalon resummations in a single formalism. A renormalon
ambiguity appears in the exponent, and the prescription necessary to
resolve it can be naturally formulated in terms of a shape function,
automatically constraining its functional form. Although the
$f_{\Tobs}$ which one gets in DGE is consistent with the one obtained
in refs. \cite{Korchemsky:2000kp,Belitsky:2001ij}, it has to be
stressed that the perturbative result $d\sigma/d\Tobs |_{pert}$ in
eq.~(\ref{EVshape}) is different in the two formalisms, since DGE
includes a class of subleading logs of renormalon origin.  In $\epem$
physics, eq.~(\ref{EVshape}) gives satisfactory results: a good fit to
the second moments of 1-thrust, $M_H$, and $C$ parameter is obtained
in ref. \cite{Korchemsky:2000kp}, and the fits for thrust and $M_H$ of
ref. \cite{Gardi:2002bg} are in better agreement in the
$(\as,\alpha_0)$ plane wrt those obtained with DMW.  On the other
hand, according to ref. \cite{Gardi:2002bg}, $\as(M_Z)=0.1086\pm
0.0004(exp)$ (the theory error is estimated to be around 5\%).
Therefore, it seems that a more refined treatment of non-perturbative
effects, which is helpful in other respects, is not what one needs in
order to get larger $\as$ values.

In a couple of interesting analyses, Delphi adopted rather unconventional
methods to study event shapes. In ref. \cite{Abreu:2000ck}, event shape
distributions were compared to fixed-order ${\cal O}(\as^2)$ (NLO) results
(i.e., resummation has not been included), using the renormalization 
scale as a free parameter in the fit, and correcting for hadronization 
effects with MC's. Although I'm not aware of any theoretical consideration
which justifies such a procedure (called in ref. \cite{Abreu:2000ck}
``experimental optimization of the scale''), the $\as$ values obtained from
different observables display a remarkable consistency, and they are
also consistent with those obtained by using NLL-resummed predictions.
In [5-228] (see also ref. \cite{Wicke:2002wh}) event shape means were
shown to give mutually consistent $\as$ values in the context of a
renormalization group approach (RGI \cite{Dhar:1983py}), without
needing any hadronization corrections. The final $\as$ results of
refs. \cite{Abreu:2000ck} and [5-228] are in excellent agreement with
the world average. I interpret these findings as the indication that
(at least in a given CM energy range) the uncertainties affecting
theoretical predictions at the NLO are larger than or of the same
order as the power-suppressed effects that one aims to study. It thus
appears that the computation of event shapes at the NNLO is necessary
for a deeper understanding of this matter.

In summary, in the past few years a solid progress has been achieved 
in the understanding of power-suppressed effects in $\epem$ collisions
and in DIS. Although models such as DMW can successfully describe 
many features of the data, some aspects deserve further studies.
In some cases, improvements are obtained within approaches which refine
the treatment of the non-perturbative part, using a shape function, but
the computation of the next order in perturbation theory will likely
be necessary in order to obtain a more consistent overall picture.
It is worth recalling that the study of hadron-mass effects has been 
found \cite{Salam:2001bd} to induce further power-suppressed terms, 
some of which can be eliminated by adopting a suitable definition for
the observables (E-scheme). Furthermore, more stringent tests of the
models for power-suppressed effects should be performed using observables 
with more complicated kinematic structure and/or gluons at the LO (see 
refs. \cite{Banfi:2001pb,Banfi:2001aq} and references therein). Finally, 
it appears to be mandatory to extend the studies of such models to the 
case of hadronic collisions (for jet observables in particular), where 
little work has been done so far.

\section{NNLO COMPUTATIONS\label{sec:NNLO}}
Bottom production at the Tevatron and event shapes in $\epem$ collisions
are a couple of examples which provide physical motivations to increase the
precision of the perturbative computations. If tree $n$-point functions
contribute to a given reaction at the LO, the N$^{k}$LO result (i.e., of 
relative order $\as^k$ wrt to the LO) will get contributions from 
the $l$-loop, $(n+p)$-point functions, with $l+p\le k$. There are
basically three major steps to make in order to get physical predictions:
{\em i}) explicit computation of all the $l$-loop, $(n+p)$-point functions;
{\em ii}) cancellation of soft and collinear divergences (which I'll
denote -- improperly -- as IR cancellation henceforth);
{\em iii}) numerical integration of the {\em finite} result obtained from 
{\em i}) and {\em ii}), with MC techniques to allow more flexibility.
One should also mention that UV renormalization has in
general to be carried out; however, this is basically textbook matter 
by now, and thus I'll not deal with it in the following. For NLO
computations ($k=1$), steps {\em i)}--{\em iii)} appear to be
understood. One-loop integrals have been computed up to five external
legs \cite{Bern:1993kr}; the case of
$n\ge 6$ cannot probably be handled with Feynman-diagram techniques
only, and still awaits for a general solution (see \cite{ZBern}). 
Subtraction \cite{Ellis:1980wv} and slicing \cite{Fabricius:1981sx}
methods, to achieve IR cancellation with semi-analytical 
techniques, have been available for a long time. In their 
modern versions \cite{Giele:1991vf,Giele:dj,Frixione:1995ms,Catani:1996jh,%
Catani:1996vz,Frixione:1997np,Nagy:1996bz} they are formulated in an 
universal (i.e., process- and $n$-independent) way, which simplifies
step {\em iii)} considerably, and allows the computation of any IR
safe observable (no matter how exclusive). No modification is needed
in order to incorporate new one-loop results. Attempts to achieve IR
cancellation through full numerical
computations \cite{Soper:1998ye,Soper:1999xk,Kramer:2002cd} are still
in a preliminary stage, reproducing known results for three-jet
production in $\epem$ collisions.

Only a handful of production processes have been computed to
NNLO accuracy and beyond: results are available for
DIS coefficient functions \cite{Zijlstra:1992qd,Zijlstra:1992kj} and 
for the Drell-Yan K-factor \cite{Hamberg:1990np,vanNeerven:1991gh} 
at ${\cal O}(\as^2)$, and for the rate $\epem\to$ 
hadrons \cite{Gorishnii:1990vf}
at ${\cal O}(\as^3)$ (I'll later deal with inclusive Higgs production 
in some details). All these computations 
are inclusive enough to allow a complete analytical
integration over the phase-space of final-state partons.
Such an integration is not possible in general, and 
an NNLO process-independent and exclusive formulation of IR 
cancellation will likely proceed through semi-analytical techniques, 
similar to those adopted at the NLO. Therefore, one
needs to know the IR-divergent pieces of all the quantities contributing
to the cross section at the NNLO. These are (I still assume that the
LO gets contribution from the tree $n$-point functions): 
{\em a)} tree-level $(n+2)$-parton amplitudes squared $\abs{T_{n+2}}^2$;
{\em b)} the interference between tree-level $(n+1)$-parton amplitudes
 and one-loop $(n+1)$-parton amplitudes $\Re{(T_{n+1}^\star L_{n+1}^{(1)})}$;
{\em c)} the interference between tree-level $n$-parton amplitudes
 and two-loop $n$-parton amplitudes $\Re{(T_n^\star L_n^{(2)})}$;
{\em d)} one-loop $n$-parton amplitudes squared $|L_n^{(1)}|^2$.
The IR divergences of $\abs{T_{n+2}}^2$ result from having two soft
partons, or three collinear partons, or one soft parton plus two other 
collinear partons, or two pairs of collinear partons. Only the latter 
configuration is trivial, in the sense that the corresponding singular
behaviour of $\abs{T_{n+2}}^2$ can be obtained from known NLO results; 
the other limits have been studied in refs. \cite{Berends:1988zn,%
Campbell:1997hg,Catani:1998nv,DelDuca:1999ha,Catani:1999ss}, and can be 
generally cast in the form of a reduced $n$-resolved-parton matrix element,
times a suitable kernel. The problem of combining these singular pieces
into local IR counterterms, and of integrating these counterterms 
over the appropriate region of the phase space, is still unsolved.
One also needs to know the IR divergences of $L_{n+1}^{(1)}$ when one
parton is soft or two partons are collinear: this is a new feature of 
NNLO computations, since at the NLO all partons in a virtual contribution 
are resolved. These limits are also known \cite{Bern:zx,Bern:1995ix,%
Bern:1998sc,Kosower:1999xi,Kosower:1999rx,Bern:1999ry,Catani:2000pi}.
Finally, the general form of the residues of the poles $1/\vep^4$, 
$1/\vep^3$, and $1/\vep^2$ appearing in $L_n^{(2)}$ has been given in
ref. \cite{Catani:1998bh}, without computing any two-loop integrals
(see also \cite{Sterman:2002qn}).

The pole terms found in ref. \cite{Catani:1998bh} must precisely match
those resulting from the explicit computations of two-loop integrals. A lot 
of progress has been made in the past couple of years in such computations, 
and now all the two-loop $2\to 2$ and $1^*\to 3$ amplitudes
are available. First, all the (very many) tensor integrals are 
reduced to a much smaller number of master scalar integrals, thanks
to integration-by-part identities \cite{Tkachov:wb,Chetyrkin:qh}
(previously used for two-point functions), and Lorentz-invariance
identities \cite{Gehrmann:1999as}; integration-by-part identities
for $n$-leg, $l$-loop integrals were also shown \cite{Baikov:2000jg} 
to be equivalent to those for $(n-m)$-leg, $(l+m)$-loop integrals.
The problem of actually computing the master integrals is of a different
nature. The breakthrough \cite{Smirnov:1999gc,Tausk:1999vh} was the
use of a Mellin-Barnes representation for the propagators in the computation
of planar and non-planar massless double box integrals (expanded in 
the dimensional-regularization parameter $\vep$). Negative
space-dimension techniques \cite{Halliday:1987an} can also be applied
to simpler topologies \cite{Anastasiou:1999cx,Suzuki:2001yf}.
The computation of master integrals is mapped onto the problem
of solving differential equations in the approach of 
refs. \cite{Kotikov:pm,Remiddi:1997ny}.
This approach has been adopted to compute all of the double box master
integrals with one off-shell leg \cite{Gehrmann:2000zt,Gehrmann:2001ck},
not all of which had been computed with Mellin-Barnes 
techniques \cite{Smirnov:2000vy,Smirnov:2000ie}.
The master integrals so far computed, together with the reduction-to-master-%
integral techniques, {\em would} allow the computation of $\epem\to~3$~jets,
and of two-jet production in hadronic collisions (and a few other processes: 
see for example ref. \cite{Giele:2002hx} for a discussion -- 
unfortunately, these processes don't include heavy flavour hadroproduction)
if one knew how to achieve IR cancellation for a generic observable
at the NNLO. 

In hadronic physics, the computation of NNLO short-distance cross sections
is not sufficient to get NNLO-accurate predictions, since NNLO-evolved PDF's
are also necessary. NNLO-evolved PDF's require the computation of 
Altarelli-Parisi splitting functions to 
three loops. It turns out to be convenient to perform such a 
computation in Mellin space; the results for the first few Mellin 
moments \cite{Larin:1993vu,Larin:1996wd,Retey:2000nq} 
(together with constraints on the small-$x$ behaviour)
have been used \cite{vanNeerven:1999ca,vanNeerven:2000uj,vanNeerven:2000wp}
to obtain {\em approximate} expressions for the splitting functions
in the $x$ space. Very recently, the complete three-loop computation 
of the $n_{\sss F}$ part of the non-singlet structure function in DIS 
has become available \cite{Moch:2002sn}, from which the corresponding
coefficient functions (relevant to N$^3$LO computations) and splitting
function can be extracted; the latter has been cross-checked 
against the approximate results mentioned above, and full agreement
has been found. Although the complete expressions for the splitting 
functions will not appear soon \cite{Moch:2002sn}, this fairly
impressive result gives confidence on the accuracy of the approximate
solution of ref. \cite{vanNeerven:2000wp}.  One has to keep in mind
that, in order for any PDF set (such as NNLO-MRST \cite{Martin:2002dr}) 
to actually be of NNLO accuracy, not only the three-loop splitting
functions are needed, but also {\em all} the short-distance cross
sections used in the fits must be computed to NNLO. At present, this
is the case for the DIS coefficient functions only (which implies that
the approximation is generally good, the bulk of the data being from
DIS); for example, in the case of Drell-Yan the $x_{\sss F}$
distribution needed for the fits has only NLO accuracy (the Drell-Yan
NNLO K-factor is used for normalization purposes \cite{Martin:2002dr}). 
This is another hint of the necessity of solving the problem of IR 
cancellation at the NNLO in a general way.

In view of enormous amount of work done and yet to be done, one may
ask whether the final outcome will justify such an effort. The answer
is certainly positive, but one must keep in mind that NNLO computations
will not automatically mean {\em precision} physics. As shown in the
case of $\epem$ event shapes, short- and large-distance effects are
always correlated, and any advancement in perturbation theory should
be complemented by a deeper understanding of this correlation.
Furthermore, no precision study in hadronic collisions can be
made without an accurate assessment of uncertainties due to PDFs.
This matter is now receiving considerable attention: see for
example ref. \cite{Giele:2002hx} for a review. 
NNLO computations will certainly play a
major role in those cases in which NLO results still give an
unsatisfactory description of data: examples are the processes
with large K-factors (such as $b$ production), or the observables
which require a better description in terms of kinematics (such
as jet profiles).

A large-K-factor process is the direct production of SM Higgs
at hadron colliders. At the NLO, the exact computation \cite{Spira:1995rr}
for the $gg$ channel is found to be in excellent agreement with approximate
results \cite{Dawson:1990zj,Djouadi:1991tk} based on keeping the 
leading term of an expansion in $\mH^2/m_{top}^2$ (the agreement 
further improves if the full dependence on $\mH^2/m_{top}^2$
is kept in the LO term). Therefore, one assumes the same to hold at
the NNLO, and computes the NNLO contribution in the $m_{top}\to\infty$
limit. This is feasible since the effective Lagrangian $ggH$ has been 
obtained \cite{Chetyrkin:1997iv,Chetyrkin:1997un} to ${\cal O}(\as^4)$
(actually, one order larger than necessary here). Exploiting the 
technique of ref. \cite{Baikov:2000jg}, the two-loop virtual correction
to $gg\to H$ has been obtained in ref. \cite{Harlander:2000mg}.
The missing contributions to the physical cross section (including
$qg$ and $q\bar{q}$ channels) have been presented in
ref. \cite{Harlander:2002wh}, where an expansion for $\xH\simeq 1$
($\xH=\mH^2/\hat{s}$) has been used to compute the double-real 
contribution. Terms up to $(1-\xH)^{16}$ have been included.
The $\xH\simeq 1$ expansion is expected to work well, since the gluon density 
is rather peaked towards small $x$'s, and thus the CM energy available at
the partonic level is never too far from threshold. As argued in
ref. \cite{Kramer:1996iq}, collinear radiation also gives a sizable
effect. Methods similar to \cite{Kramer:1996iq} have been used in
refs. \cite{Catani:2001ic,Harlander:2001is}, with the result of 
ref. \cite{Harlander:2000mg}, to give early estimates of the complete
NNLO rate. These estimates are seen to agree well with the result of 
ref. \cite{Harlander:2002wh}, which confirms the 
soft-collinear dominance in inclusive Higgs production at colliders.
Finally, the result of ref. \cite{Harlander:2002wh} has been found to 
agree to 1\% level or better with the computation of 
ref. \cite{Anastasiou:2002yz}, where the double-real contribution
is also evaluated exactly (it is interesting to notice
that in ref. \cite{Anastasiou:2002yz} the phase-space integrals
relevant to real-emission terms are computed with techniques used so far
only for loop diagrams -- it remains to be seen whether this method can
be generalized to more exclusive observables).
\begin{figure}[htb]
\includegraphics[width=15pc]
{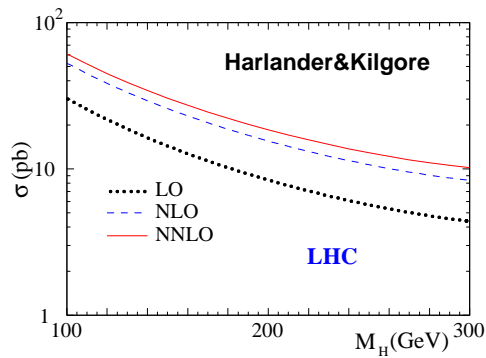}
\vskip -2.2pc
\caption{SM Higgs total rate at the LHC.}
\label{fig:HiggsatLHC}
\end{figure}
As shown in fig.~\ref{fig:HiggsatLHC}, the inclusion of NNLO
corrections seems to suggest that effects beyond this order are 
negligible. The scale dependence is reduced wrt the one observed
at the NLO (see ref. \cite{Anastasiou:2002yz}). The dominance of
the region $\xH\sim 1$ implies the potential relevance of soft-gluon
resummation. Preliminary results \cite{Grazzini:2002jm,Giele:2002hx}
indicate that NNLL resummation enhances the NNLO rate by 5--6\%
(12--15\%) at the LHC (Tevatron), for $100<\mH<200$~GeV.

The result for the fully-inclusive rate also serves to compute a
slightly less inclusive observable, namely the rate for Higgs+jets,
with the $\pt$ of any jets imposed to be smaller than a fixed quantity
$p_{\sss T}^{(veto)}$ (jet veto). Such an observable, which should
help in reducing the background due to the decay channel $H\to
W^*W^*$, has been computed in ref. \cite{Catani:2001cr} by subtracting
the anti-vetoed jet cross section $\pt>p_{\sss T}^{(veto)}$ (obtained
in ref. \cite{deFlorian:1999zd}) from the inclusive NNLO result
discussed so far. The study of more exclusive observables, which
implies the understanding of IR cancellation at NNLO, will certainly
prove useful in the future. In this case, the dominance of the region
$\xH\sim 1$ might not be as strong as in the case of inclusive rates.
Although it is unlikely that the region $\xH\sim 0$ will play any role
in phenomenological studies, it is worth recalling that in this region
the approximation $m_{top}\to\infty$ is not expected to work well: 
in the full theory the dominant contribution for $\xH\to 0$ is
single-logarithmic, whereas double logs are also found in the
large-$m_{top}$ theory.  The latter terms have been identified
explicitly in ref. \cite{Hautmann:2002tu} with $\kt$-factorization
arguments; at the NNLO, they are seen to coincide with those resulting
from the explicit computation of ref. \cite{Anastasiou:2002yz}, thus
providing a cross check impossible to achieve in the comparison of ref
\cite{Anastasiou:2002yz} with ref. \cite{Harlander:2002wh}.

\section{MONTE CARLO SIMULATIONS}
Monte Carlo (MC) programs are essential tools in experimental physics,
giving fully-fledged descriptions of hadronic final states which
cannot be obtained in fixed-order computations. Schematically, an MC
works as follows: for a given process, which at the LO receives
contribution from $2\to n_0$ reactions, $(2+n_0)$-particle
configurations are generated, according to exact tree-level matrix
element (ME) computations. The quarks and gluons (partons henceforth)
among these primary particles are then allowed to emit more quarks and
gluons, which are obtained from a parton shower or dipole cascade 
{\em approximation} to QCD dynamics. This implies that MC's cannot
simulate the emission of final-state hard (i.e., with large relative
transverse momenta; thus, hard is synonymous of resolved here) partons
other than the primary ones obtained from ME computations.
Furthermore, total rates are accurate to LO.

Although these problems are always present in MC simulations, they
become acute when CM energies grow large, since in this case channels
with large numbers of well-separated jets are
phenomenologically very important and, correspondingly, total rates
need to be computed to an accuracy better than LO. Two strategies can
be devised in order to improve MC's. The first aims at having $\nE$
extra hard partons in the final state; thus, in the example given
above, the number of final-state hard particles would increase from
$n_0$ to $n_0+\nE$. This approach is usually referred to as 
{\em matrix element corrections}, since the MC must use the
$(2+n_0+\nE)$-particle ME's to generate the correct hard kinematics;
more details are given in sect.~\ref{sec:MEcorr}. The second strategy
also aims at simulating the production of $n_0+\nE$ hard particles,
but improves the computation of rates as well, to N$^{\nE}$LO
accuracy. A discussion is given in sect.~\ref{sec:MCatNLO}.

\subsection{Matrix element corrections\label{sec:MEcorr}}
There are basically two major problems in the implementation of ME
corrections. The first problem is that of achieving a fast computation
of the ME's themselves for the largest possible $n_0+\nE$, and an
efficient phase-space generation.  The second problem stems from the
fact that multi-parton ME's are IR divergent. Clearly, in
hard-particle configurations IR divergences don't appear; however, the
definition of what hard means is, to a large extent, arbitrary. In
practice, hardness is achieved by imposing some cuts on suitable
partonic variables, such as $\pt$'s and $(\eta,\varphi)$-distances
$dR$ in hadronic collisions.  I collectively denote these cuts by
$\delta_{sep}$. One assumes that $n$ hard partons will result (after
the shower) into $n$ jets; but, with a probability depending on
$\delta_{sep}$, a given $n$-jet event could also result from
$n+m$ hard partons. This means that, when generating
events at a fixed $n_0+\nE$ number of primary particles, physical
observables in general depend upon $\delta_{sep}$; I refer to this as
the $\delta_{sep}$-bias problem. Any solution to the
$\delta_{sep}$-bias problem implies a procedure to combine
consistently the treatment of ME's with different $n_0+\nE$'s. Here,
the difficulty is that of avoiding double counting, that is, the
generation of the same kinematical configuration more often than
prescribed by QCD.

The vast majority of recent approaches to ME corrections address 
only the first of the two problems mentioned above. A considerable
amount of work has been devoted to the coding of hadronic processes with 
vector bosons/Higgs plus heavy quarks in the final state, which cannot
be found (regardless of the number of extra partons) in standard MC's.
The complexity of hard-process generation for large $n_0+\nE$ suggests to
implement it in a package (which I call ME generator) distinct from
the shower MC. The ME generator stores a set of hard configurations
in a file (event file); the event file is eventually read by the
MC, which uses the hard configurations as initial conditions for the
showers.  The advantage of this procedure is that it is completely
modular: one given event file can be read by different MC's, and
conversely one MC can read event files produced by different ME
generators. It is clearly convenient to reach an agreement on the 
format of such event records: this is now available \cite{Boos:2001cv}
(Les Houches accord \#1). Ready-to-use ME generators (with different
numbers of hard processes implemented) are AcerMC \cite{Kersevan:2002dd}, 
ALPGEN \cite{Caravaglios:1998yr,Mangano:2001xp,Mangano:2002ea}, and
MadGraph/MadEvent \cite{Stelzer:1994ta,Maltoni:2002qb}.
Related work, at present set up to function only with Pythia, has 
been presented by the CompHEP \cite{Pukhov:1999gg,Belyaev:2000wn} 
and Grace \cite{Sato:2001ae,Tsuno:2002ce} groups.
All of these ME generators use Feynman-diagram techniques in the 
computation of ME's, except ALPGEN, which uses the iterative algorithm 
Alpha \cite{Caravaglios:1995cd} (see also \cite{Draggiotis:2002hm}).

\begin{figure}[htb]
\includegraphics[width=15pc]{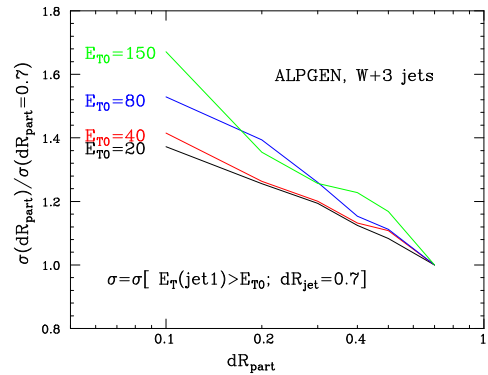}
\vskip -2.2pc
\caption{Dependence of (normalized) jet rates on interparton separation.}
\label{fig:WjetALPGEN}
\end{figure}

When using an ME generator, the cuts $\delta_{sep}$ must be looser
than those used to define the observables. For example, the $\pt$-cut
imposed at the parton level must be smaller than the minimum $\pt$ of
any jets. On the other hand, the cuts should not be too loose: the
looser the cuts, the larger the probability of getting a $n$-jet event
starting from $n+m$ hard partons. Thus, $\delta_{sep}$ must
be chosen in a range which is somewhat dependent upon the observables
that one wants to study. This implies that, strictly speaking, the
combination of an ME generator with a shower MC is {\em not} an event
generator, since the event record depends upon the observables. This
happens precisely because such a combination is affected by the
$\delta_{sep}$-bias problem: it is therefore necessary to assess
its impact on physical observables.
An example is given in fig.~\ref{fig:WjetALPGEN}, obtained using 
ALPGEN+Herwig (any other ME generator and shower MC would give
equivalent results for the same observable). The plot presents jet 
rates, integrated over $E_{\sss T}^{(jet)}>E_{{\sss T}0}$, 
for the hardest jet in W+3-jet events at the Tevatron, versus
the parton separation $dR_{part}$ imposed at the level of ME
generation. Jets are reconstructed with the cone algorithm, with
$R=0.7$. Rates are normalized to the result obtained with
$dR_{part}=0.7$. The conclusion is that, in the ``reasonable'' range
$0.3<dR_{part}<0.7$, the physical prediction has a ${\cal O}(20\%)$
dependence on $dR_{part}$. The existence of a $\delta_{sep}$
dependence should always be kept in mind when using an ME generator +
shower MC combination, because it affects the precision of the
prediction. On the other hand, this is a rather modest price to pay:
for multi-jet observables, ordinary MC's can underestimate the cross
section by orders of magnitude, and the use of ME generators is
mandatory.

As mentioned before, the $\delta_{sep}$ bias can be avoided by
suitably combining the generation of ME's with different $n_0+\nE$.
Early proposal for ME corrections \cite{Gustafson:1986db,Bengtsson:1986hr,%
Bengtsson:1986et,Gustafson:1987rq,Seymour:1994df} achieved this in
the case $\nE=0,1$. The solution for arbitrary $\nE$ appears to be
more complicated; it has been fully implemented for shower MC's in
$\epem$ collisions \cite{Catani:2001cc,Kuhn:2000dk} in the case of jet
production; along similar lines, proposal for colour dipole MC's
\cite{Lonnblad:2001iq} and shower MC's in hadronic collisions
\cite{Krauss:2002up} have also been made. The idea of ref.
\cite{Catani:2001cc} is the following. {\em a)} Integrate all the
$\gamma^*\to 2+\nE$ ME's by imposing $y_{ij}>y_{\sss INI}$ for any
pairs of partons $i,j$, with $y_{\sss INI}$ a fixed parameter and
$y_{ij}=2\min(E_i^2,E_j^2)(1-\cos\theta_{ij})/Q^2$ the interparton
distance defined according to the $\kt$-algorithm
\cite{Catani:1991hj}.  {\em b)} Choose statistically an $\nE$, using
the rates computed in {\em a)}.  {\em c)} Generate a $(2+\nE)$-parton
configuration using the exact $\gamma^*\to 2+\nE$ ME, and reweight it
with a suitable combination of Sudakov form factors (corresponding to
the probability of no other branchings). {\em d)} Use the configuration 
generated in {\em c)} as initial condition for a {\em vetoed} shower.  
A vetoed shower proceeds as the usual one, except that it forbids all
branchings $i\to jk$ with $y_{jk}>y_{\sss INI}$ without stopping the
scale evolution. In ref \cite{Catani:2001cc}, $y_{\sss INI}$ plays the
role of $\delta_{sep}$.  Although the selection of an $\nE$ value has
a leading-log dependence on $y_{\sss INI}$, it can be proved that this
dependence is cancelled up to next-to-next-to-leading logs in physical
observables \cite{Catani:2001cc}.
\begin{figure}[htb]
\includegraphics[width=15pc]{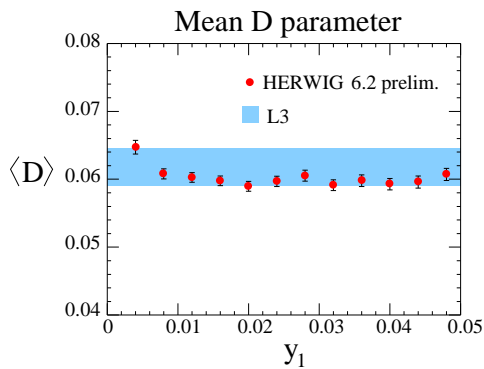}
\vskip -2.2pc
\caption{Dependence of mean $D$ parameter on $y_1\equiv y_{\sss INI}$.}
\label{fig:meanD}
\end{figure}
This is illustrated in fig.~\ref{fig:meanD}, where the mean value of
the $D$ parameter \cite{Parisi:1978eg,Ellis:1980wv} (full circles) is
plotted versus the value of $y_{\sss INI}$, and compared to the
measurement of ref. \cite{Adeva:1992gv} (band). Figure~\ref{fig:meanD}
clearly documents that the consistent combination of ME's with
different $n_0+\nE$ largely reduces the dependence of observables on
$\delta_{sep}$. This, however, comes at the price of modifying the
shower algorithm.  Furthermore, the procedure is more
computing-intensive, since {\em all} $\nE$'s need to be considered
(this being impossible in practice, in $\epem$ collisions the
procedure has an error of ${\cal O}(\as^{N-1})$ if only ME's with
$2+\nE\le N$ partons are considered; in ref. \cite{Kuhn:2000dk},
$N=5$).

\subsection{Complete matching of Monte Carlos and perturbative computations
\label{sec:MCatNLO}}
The problem of fully matching MC's with higher-order computations can
be seen as an upgrade of ME corrections: not only one wants to
describe the kinematics of $n_0+\nE$ hard particles correctly, but the
information on N$^{\nE}$LO rates must also be included. First attempts
at solving this problem have only recently become available, and only
for the case $\nE=1$. Following ref. \cite{Frixione:2002ik}, let me
denote by $\MCatNLO$ the improved MC we aim at constructing. The naive
idea, of defining an $\MCatNLO$ by multiplying an MC with ME
corrections by the NLO K-factor, is simply not acceptable: the
inclusiveness of the K-factor does not really fit well into the
exclusive framework of an MC. Thus, any $\MCatNLO$ must involve the
computation of virtual ME's. This is the reason why the construction
of an $\MCatNLO$ is conceptually more complicated than ME corrections:
the IR divergences of the virtual ME's can only be cancelled by
computing real-emission ME's with one soft parton or two collinear partons.
These soft and collinear configurations never occur in ordinary MC's;
in the case of ME corrections, the cuts $\delta_{sep}$ are
specifically introduced to avoid them. The presence of virtual ME's
also requires a less-intuitive definition of double counting
\cite{Frixione:2002ik}: in the context of $\MCatNLO$'s, double
counting may correspond to either an excess or a deficit in the
prediction. This generalization is necessary, since the ME's used in
$\MCatNLO$'s are not positive-definite.

In order to describe in more details current approaches to $\MCatNLO$,
I adopt the toy model of ref. \cite{Frixione:2002ik}, in which a 
system $S$ (say, a quark) can emit ``photons'', massless particles
with only one degree of freedom (say, the energy). The initial energy
of $S$ is 1, which becomes $1-x$ after the emission of one photon 
of energy $0<x\le 1$. The LO, virtual, and real contributions to
the NLO cross section are ($a$ is the coupling constant):
\beqn
\xsborn&=&B\delta(x),
\label{born}
\\
\xsvirt&=&a\left(\frac{B}{2\vep}+V\right)\delta(x),
\label{virt}
\\
\xsreal&=&a\frac{R(x)}{x},
\label{real}
\eeqn
where $\delta(x)$ in eqs.~(\ref{born}) and~(\ref{virt}) reminds 
that there's no emission of real photons, and the IR divergence 
$1/\vep$ in eq.~(\ref{virt}) results from the loop integration
over virtual photon momentum in $4-2\vep$ dimensions. 
I denote by $(S,z)$ the configuration of the system plus up to
one photon, with $z=0$ in the case of the LO or virtual 
contributions (eqs.~(\ref{born}) and ~(\ref{virt}) respectively),
and $z=x\ne 0$ in the case of the real contribution (eq.~(\ref{real})).
Energy conservation is understood, and therefore in the 
configuration $(S,z)$ the system has energy $1-z$. 
The function $R(x)$ characterizes real emissions; its specific
form is irrelevant, except that, for IR cancellation to occur, it
must fulfil $R(x)\to B$ for $x\to 0$. With one photon emission at
most, any observable $O$ can be represented by a function $O(S,z)$; the
computation of its expectation value $\VEV{O}$ can be achieved through
standard techniques for IR cancellation:
\beqn
\VEV{O}&=&BO(S,0)+a\Bigg[\left(B\log\delta +V\right)O(S,0) 
\nonumber \\*&+& 
\int_\delta^1 dx\, \frac{O(S,x)R(x)}{x}\Bigg]
\label{nloslicing}
\eeqn
in the slicing method \cite{Fabricius:1981sx}, and
\beqn
\VEV{O}&=&\int_0^1 dx \Bigg[O(S,x)\frac{aR(x)}{x}
\nonumber \\*&+& 
O(S,0)\left(B+aV-\frac{aB}{x}\right)\Bigg]
\label{nlosubtint}
\eeqn
in the subtraction method \cite{Ellis:1980wv}. 

In an MC approach, the system can undergo an arbitrary 
number of photon emissions. I denote by $B\IMC(O;S,0)$ the distribution
in the observable $O$ obtained with MC methods; this notation reminds
that in a standard MC the initial condition for the shower is
$(S,0)$ (the LO kinematics), and that the total rate is $B$ (the LO 
rate, see eq.~(\ref{born})). The most straightforward implementation
of an $\MCatNLO$ can then be done by analogy: since two kinematical
configurations, $(S,x)$ and $(S,0)$, appear in the NLO cross section,
one can use both of them as initial conditions for the showers.
In order to recover the correct total rate, each event resulting from
a shower with initial condition $(S,z)$ will be weighted
with the coefficient of $O(S,z)$ which appears in 
eq.~(\ref{nloslicing}) or in eq.~(\ref{nlosubtint}). Using 
eq.~(\ref{nlosubtint}), one gets
\beqn
\left(\xsecO\right)
&=&\int_0^1 dx \Bigg[\IMC(O;S,x)\frac{aR(x)}{x}
\nonumber\\*&+&
\IMC(O;S,0)\left(B+aV-\frac{aB}{x}\right)\Bigg].\phantom{aaaa}
\label{IMCnlonaive}
\eeqn
Unfortunately, this naive approach does not work. The weights
$aR(x)/x$ and $B+aV-aB/x$ are IR-divergent at $x=0$; since the
corresponding showers have different initial conditions $(S,x)$ and
$(S,0)$, it would take an infinite amount of time to cancel the
divergences (in other words, unweighting is impossible). This is not a
practical problem, is a fundamental one: the cancellation works for
inclusive quantities, and the shower is exclusive. So the main problem
in the construction of an $\MCatNLO$ can be reformulated as follows:
how to achieve IR cancellation, without giving up the exclusive
properties of the showers. Besides, eq.~(\ref{IMCnlonaive}) also
suffers from double counting.

In the context of the slicing method, an approach has been proposed
\cite{Potter:2001an,Potter:2001ej,Dobbs:2001gb,Dobbs:2001dq} which
exploits an idea of refs. \cite{Baer:1991qf,Baer:1991ca} (see also
\cite{Baer:1996vx}). The slicing parameter $\delta$ in
eq.~(\ref{nloslicing}) is fixed to the value $\delta_0$, by imposing
that no $(S,0)$ contribution be present in the NLO cross section:
\beq
B+a\left(B\log\delta_0 +V\right)=0\,.
\label{getdeltaz}
\eeq
This effectively restricts the energy of the real photons emitted to
the range $\delta_0<x\le 1$ (see eq.~(\ref{nloslicing})). This range is
further partitioned by means of an arbitrary parameter $\delta_{\sss PS}$.
One starts by generating the emission of a real photon with energy $x$
distributed according to $aR(x)/x$. Then, if $\delta_0<x\le\delta_{\sss PS}$,
the real-emission kinematics $(S,x)$ is mapped onto the LO kinematics
$(S,0)$ (in other words, the photon with energy $x$ is thrown away).
The configuration $(S,0)$ is used as initial condition for the shower,
requiring the shower to forbid photon energies larger than $\delta_{\sss PS}$. 
If $x>\delta_{\sss PS}$, the real emission is kept, and $(S,x)$ is used 
as initial condition for the shower. The corresponding formula is:
\beqn
&&\!\!\!\!\!\!\!\!\!\!\!\!\left(\xsecO\right)\!=\!a\int_0^1 dx\Bigg[
\IMC(O;S,x)\frac{R(x)}{x} \stepf(x-\delta_{\sss PS})
\nonumber \\*&&+
\IMC(O;S,0)\frac{R(x)}{x} 
\stepf(x-\delta_0)\stepf(\delta_{\sss PS}-x)\Bigg].\phantom{aaaa}
\label{PhiVMC}
\eeqn
The advantage of eq.~(\ref{PhiVMC}) is that it is manifestly
positive-definite, and that its implementation can be carried out with
little or no knowledge of the structure of the MC.  On the other hand,
it can be shown \cite{Frixione:2002ik} that eq.~(\ref{PhiVMC}) still
has double counting, and does not have a formal perturbative expansion
in $a$. These problems arise since the technique adopted to deal with
the IR cancellation is not exclusive enough: eq.~(\ref{getdeltaz}) is an
integral equation (the $\log\delta_0$ term is due to an integral over
soft-photon configurations in the real-emission contribution).
Although the absence of a perturbative expansion can be seen as a
minor drawback from a practical point of view, the impact of the
double counting should be assessed for each observable studied, by
considering the dependence of physical predictions upon $\delta_{\sss PS}$.

Other approaches \cite{Frixione:2002ik,Collins:2000qd,Chen:2001ci,%
Collins:2001fm,Chen:2001nf,Kurihara:2002mv} are based on the 
subtraction method. One can observe that ordinary MC's do
contain the information on the leading IR singular behaviour of
NLO ME's. Formally, this implies that the ${\cal O}(\as)$
term in the perturbative expansion of the MC result can act as a 
{\em local} counterterm to the IR divergences at the NLO (strictly
speaking, this is not exactly true in the case of large-angle soft
gluon emission, and a few technical complications arise -- see
ref. \cite{Frixione:2002ik}). Furthermore, the form of the counterterm
does not depend upon the observable studied. Thus, IR cancellation
is achieved locally, but without any reference to a specific observable:
this allows to implement it at the level of event generation, without
giving up the exclusive treatment of the branchings. The prescription 
of ref. \cite{Frixione:2002ik} is
\beqn
&&\!\!\!\!\!\!\!\!\!\!\!\!\left(\xsecO\right)
\!=\!\int_0^1 dx \Bigg[\IMC(O;S,x)\frac{a[R(x)-BQ(x)]}{x}
\nonumber \\*&&+
\IMC(O;S,0)\left(B+aV+\frac{aB[Q(x)-1]}{x}\right)\Bigg],\phantom{aaaa}
\label{IMCfive}
\eeqn
where the $Q(x)$-dependent quantities are the ${\cal O}(a)$ term of the
MC result. The local IR cancellation mentioned before shows in the 
fact that the coefficients of $\IMC(O;S,x)$ and $\IMC(O;S,0)$ in 
eq.~(\ref{IMCfive}) are finite, since the condition $Q(x)\to 1$ for 
$x\to 0$ always holds, regardless of the specific MC used.
Therefore, these coefficients can be given as weights
to the showers with $(S,x)$ and $(S,0)$ initial conditions respectively.
Eq.~(\ref{IMCfive}) does not have double counting, and 
features a smooth matching between the soft- and hard-emission
regions of the phase space, without the need to introduce any 
extra parameter such as $\delta_{\sss PS}$. The price to pay for this is
the presence of negative weights (which however do not spoil the
probabilistic interpretation of the results). Furthermore, one needs
to know details of the MC ($Q(x)$) in order to implement
eq.~(\ref{IMCfive}): this seems unavoidable, since one should
expect different MC's to match differently with a given NLO computation.
The first QCD implementation of eq.~(\ref{IMCfive}) has been presented
in ref. \cite{Frixione:2002bd}.

The approach of 
refs. \cite{Collins:2000qd,Chen:2001ci,Collins:2001fm,Chen:2001nf} 
uses a technique similar to that of ref. \cite{Frixione:2002ik},
based on the definition of a local IR counterterm. At variance with
ref. \cite{Frixione:2002ik}, where the resummation of large logs
is performed to LL accuracy, 
refs. \cite{Collins:2000qd,Chen:2001ci,Collins:2001fm,Chen:2001nf} 
advocate an NLL (or beyond) resummation; for this to happen, it is argued 
that the standard formulation of collinear factorization must be extended.
This approach has been fully formulated \cite{Collins:2001fm} only in
the unphysical $\phi^3_{d=6}$ theory so far. Current QCD implementations
do not include gluon emission.

\section{CONCLUDING REMARKS}
It seems appropriate to start this section by mentioning a couple of
phenomenological issues which would have deserved more attention. One
is the problem affecting the single-inclusive, isolated-photon
measurements at the Tevatron. If D0 data are considered
\cite{Abbott:1999kd,Abazov:2001af}, a moderate disagreement with NLO
QCD is present in the low-$\pt$, central-$\eta$ region, which
disappears when the ratio $R=\sigma(\sqrt{S}=1800~{\rm GeV})/
\sigma(\sqrt{S}=630~{\rm GeV})$ is considered. The situation
worsens in the case of CDF data \cite{Acosta:2002ya}: not only the
discrepancy with QCD is statistically more significant for cross
sections, but the measured ratio $R$ also disagrees with theory. The
consistent inclusion of recoil effects, along the lines of refs.
\cite{Laenen:2000ij,Kulesza:2002rh}, might increase the QCD prediction
at small $\pt$'s, and thus reduce the discrepancy.  The second problem
affects the single-inclusive jet cross section as measured by D0
\cite{Abazov:2001hb}, the jets being reconstructed with a
$\kt$-algorithm. The data display a rather poor agreement with theory
for $\pt<100$~GeV; this is disappointing, given the excellent results
obtained with the cone algorithm (as far as the cone algorithm is
concerned, I should also mention here that the previously reported
excess of data over theory at large $\pt$ has now completely
disappeared: NLO QCD perfectly reproduces the data, if updated PDF
sets are used.  An explanation in terms of PDFs has been already given
in the past, but the PDF set used at that time resulted from an 
{\em ad hoc} fit -- named ``HJ'' by the CTEQ collaboration. This is not
necessary any longer, since in the newest PDF releases the gluon
density of the {\em best} fit naturally results to be HJ-like. See
ref. \cite{Pumplin:2002vw} for a discussion on this point). It is hard
to believe that the discrepancy in the case of the $\kt$-algorithm is
the signal of a serious problem in QCD (since it would probably affect
the cone algorithm as well); however, it may indicate a deficiency in
current MC simulations, or in the understanding of hadronization
and/or detector effects, which would surely worsen at the LHC energies. 
It has to be remarked that the $\kt$-algorithm appears to work well at HERA.

In general, the capability of QCD to describe hard production processes
is quite remarkable (with the exception of $b$-production at LEP and 
HERA). The definition of a formalism for the computation of 
exclusive observables at the NNLO is one of the most challenging
and hot topics in perturbative QCD, which will have an important 
impact on phenomenological studies and will be crucial in improving 
the precision of $\as$ (and other fundamental parameters) measurements
-- at colliders, exclusive Drell-Yan production will necessarily have 
to be calculated to NNLO, in order to match the experimental precision.
The substantial progress made in the past couple of years in the
computation of two-loop integrals and three-loop splitting functions,
and the NNLO results for direct inclusive SM-Higgs production,
are certainly very encouraging. NNLO results will also help to
understand better the interplay between soft and hard physics;
the $\epem$ and DIS environments will serve as a laboratory for the more
involved case of hadronic collisions, where most of the work remains
to be done. Monte Carlos are much better equipped to face the
challenges of Tevatron Run II and LHC. The implementation
of new processes in matrix element generators proceeds steadily and,
although there is still work to be done on the general structure of 
matrix element corrections, these techniques are rather well established 
by now. Formalisms for $\MCatNLO$'s have received considerable attention 
in the recent past; the field has still to reach a mature stage, and
new ideas will certainly be presented in the near future. A rather
obvious but quite challenging step is that of merging matrix element
corrections and $\MCatNLO$ approaches.

It is a pleasure to thank my friends and colleagues, unfortunately
too numerous to mention, who helped me so much in the preparation 
of the talk and of this manuscript.

\end{document}